\documentclass[12pt,onecolumn]{revtex4}
\usepackage{amssymb}

\begin{document}

\title{Accelerated expansion of the Universe in Gauss-Bonnet gravity}
\author{M. H. Dehghani}\email{mhd@shirazu.ac.ir}
\address{Physics Department and Biruni  Observatory,
         Shiraz University, Shiraz 71454, Iran\\ and\\
         Institute for Studies in Theoretical Physics and Mathematics (IPM)\\
         P.O. Box 19395-5531, Tehran, Iran}
\begin{abstract}
We show that in Gauss-Bonnet gravity with negative Gauss-Bonnet
coefficient and without a cosmological constant, one can explain
the acceleration of the expanding Universe. We first introduce a
solution of the Gauss-Bonnet gravity with negative Gauss-Bonnet
coefficient and no cosmological constant term in an empty
$(n+1)$-dimensional bulk. This solution can generate a de Sitter
spacetime with curvature $n(n+1)/\{(n-2)(n-3)|\alpha|\}$. We show
that an $(n-1)$-dimensional brane embedded in this bulk can have
an expanding feature with acceleration. We also considered a
4-dimensional brane world in a 5-dimensional empty space with zero
cosmological constant and obtain the modified Friedmann equations.
The solution of these modified equations in matter-dominated era
presents an expanding Universe with negative deceleration and
positive jerk which is consistent with the recent cosmological
data. We also find that for this solution, the $``n"th$ derivative
of the scale factor with respect to time can be expressed only in
terms of Hubble and deceleration parameters.
\end{abstract}

\maketitle


\section{Introduction}

The idea of brane cosmology has attracted a great deal of attention
recently. It may provide us a new solution for the so-called hierarchy
problem or the source of dark energy and dark matter \cite{Luk,Bin}. This
theory is also consistent with string theory, which suggests that matter and
gauge interaction (described by an open string) may be localized on a brane,
embedded into a higher dimensional spacetime, while the field represented by
closed strings, in particular gravity, propagate in the whole of spacetime.
In the brane world scenario, we live in a 4-dimensional (4-D) hypersurface
embedded in a higher dimensional bulk spacetime.

Among many brane models, those proposed by Randall and Sundrum which are
motivated by superstring/M-theory \cite{Wit1} are the most applicable ones.
In their first model (RS I) \cite{Ran1}, they proposed a mechanism to solve
the hierarchy problem with two branes, while in their second model (RS II)
\cite{Ran2}, they consider a single brane with a positive tension, where 4-D
Newtonian gravity is recovered at low energies even if the extra dimension
is not compact.

Brane world cosmology underscores the need to consider gravity in
higher dimensions. In this context one may use another consistent
theory of gravity in any dimension with a more general action.
This action may be written, for example, through the use of string
theory. The effect of string theory on classical gravitational
physics is usually investigated by means of a low energy effective
action which describes gravity at the classical level \cite{Wit2}.
This effective action consists of the Einstein-Hilbert action plus
curvature-squared terms and higher powers as well, and in general
gives rise to fourth order field equations and bring in ghosts.
However, if the effective action contains the higher powers of
curvature in particular combinations, then only second order field
equations are produced and consequently no ghosts arise
\cite{Zum}. The effective action obtained by this argument is
precisely of the form proposed by Lovelock \cite{Lov}. The
appearance of higher derivative gravitational terms can be seen
also in the renormalization of quantum field theory in curved
spacetime \cite{BDav}.

These facts provide a strong motivation for investigating how the
usual features of brane world cosmology are modified by more
general gravitational theories such as Lovelock gravity. In this
context, many authors are extensively studying the brane world
cosmology in Gauss-Bonnet gravity \cite {Odin}. In these analysis,
one needs the solutions of Gauss-Bonnet gravity in the bulk
spacetime. Because of the nonlinearity of the field equations, it
is very difficult to find out nontrivial exact analytical
solutions of Einstein's equation with higher curvature terms.
However, a few exact solutions of this theory have been obtained
\cite{Des,Deh}.

On the other hand, it seems established that at the present epoch the
Universe expands with acceleration instead of deceleration along the scheme
of standard Friedmann model. This follows directly from the observation of
high red-shift supernova \cite{Per} and indirectly from the measurement of
angular fluctuations of cosmic microwave background fluctuations \cite{Lee}.
The latter shows that the total mass/energy density of the Universe is very
close to critical value one ($\Omega =1$), while the observations of the
Universe in large scale structure indicate that normal gravitating (visible
and invisible) matter can contribute only $30\%$ into the total one. Thus,
one concludes that the remaining $70\%$ is some mysterious agent that
creates the cosmological acceleration. The simplest suggestion is that the
source of this acceleration is the vacuum energy (cosmological constant).
However, it meets the two well known cosmological and coincidence problems
\cite{Cal}. A second approach is the consideration of cosmological
components as a dynamical term. This scheme, usually called quintessence,
can be achieved by adding a scalar field into Einstein gravity. Several
forms of potential achieve quintessence prescriptions but non of them seems
to be directly related to some fundamental quantum field theory. A third
approach is to obtain a comprehensive model derived from some effective
theory of quantum gravity which through an inflationary period results into
the today accelerated expansion of the Universe. From this point of view,
gravitational theories including higher order curvature terms naturally come
into the game \cite{hdg}. However, some of these modified gravitational
theory have their own problems \cite{Dol}.

Here we want to explain accelerating Universe by use of the Gauss-Bonnet
gravity without a cosmological constant term or a scalar field or any other
kind of dark energy. Thus, we restrict ourself to the second and third terms
of Lovelock gravity. The second term is the Einstein-Hilbert term, while the
third term is known as the Gauss-Bonnet term. From a geometric point of
view, the combination of the Gauss-Bonnet terms constitutes, for
five-dimensional spacetimes, the most general Lagrangian producing second
order field equations, as in the four-dimensional gravity where the
Einstein-Hilbert action is the most general Lagrangian producing second
order field equations \cite{Lan}.

The outline of our paper is as follows. We give a brief review of the field
equations in Sec. \ref{Fiel}. In Sec. \ref{dS} we introduce a solution of
the Gauss-Bonnet gravity with negative Gauss-Bonnet coefficient and no
cosmological constant in an empty $(n+1)$-dimensional bulk, and show that
this solution can generate a de Sitter spacetime. In Sec. \ref{Fried} we
first introduce the modified Friedmann equations, and then obtain the
solutions of these equation in matter-dominated era, which can discuss today
accelerated expansion of the Universe. We finish our paper with some
concluding remarks.

\section{Field Equations in Gauss-Bonnet Gravity}

\label{Fiel}

The most fundamental assumption in standard general relativity is the
requirement that the field equations be generally covariant and contain at
most second order derivative of the metric. Based on this principle, the
most general Lagrangian in arbitrary dimensions is the Lovelock Lagrangian.
The Lagrangian of the Lovelock theory, which is the sum of dimensionally
extended Euler densities, may be written as
\begin{equation}
\mathcal{L}_{G}=\frac{1}{2}\sum_{i=1}^{[n/2]}c_{i}\mathcal{L}_{i},
\label{Lov1}
\end{equation}
where $c_{i}$ is an arbitrary constant, $[x]$ denotes the integer part of $x$%
, and $\mathcal{L}_{i}$ is the Euler density of a $2i$-dimensional manifold,
\begin{equation}
\mathcal{L}_{i}=(-2)^{-i}
\delta_{c_{1}d_{1}...c_{i}d_{i}}^{a_{1}b_{1}...a_{i}b_{i}}\mathcal{R}_{a_{1}
b_{1}}^{\ \ \ c_{1} d_{1}}....\mathcal{R}_{a_{i} b_{i}}^{\ \ \ c_{i} d_{i}}.
\label{Lov2}
\end{equation}
In Eq. (\ref{Lov2}) $\mathcal{R}_{a b c d}$ is the Riemann tensor and $%
\delta_{c_{1}d_{1}...c_{i}d_{i}}^{a_{1}b_{1}...a_{i}b_{i}}$ is the totally
antisymmetric product of $i$ Kronecker deltas, normalized to values 0 and $%
\pm1$. Here the first term $c_{1}\mathcal{L}_{1}=\mathcal{R}$, is just the
Einstein-Hilbert, and the second term $c_{2}\mathcal{L}_{2}=\alpha (\mathcal{%
R}_{\mu \nu \gamma \delta }\mathcal{R}^{\mu \nu \gamma \delta }-4\mathcal{R}%
_{\mu \nu }R^{\mu \nu }+\mathcal{R}^{2})$ gives us the Gauss-Bonnet term. Of
course, one may add a constant term to the above Lagrangian, playing the
role of cosmological constant term. But, as we mentioned before this creates
its own problems and therefore we don't disturb ourself with it. The
Lagrangian of Gauss-Bonnet gravity with zero cosmological constant is
\begin{equation}
\mathcal{L}_{G}=\frac{1}{2}\{\mathcal{R}+\alpha (\mathcal{R}_{\mu \nu \gamma
\delta }\mathcal{R}^{\mu \nu \gamma \delta }-4\mathcal{R}_{\mu \nu }\mathcal{%
R}^{\mu \nu }+\mathcal{R}^{2})\},  \label{Actg}
\end{equation}
where $\mathcal{R}$, $\mathcal{R}_{\mu \nu \rho \sigma }$, and $\mathcal{R}%
_{\mu \nu }$ are the Ricci scalar and Riemann and Ricci tensors of the
spacetime, and $\alpha $ is the Gauss-Bonnet coefficient with dimension $(%
\text{length})^{2}$. Here we restrict ourselves to the case $\alpha < 0$.

Let us suppose, as would be in the co-dimension one brane world scenario,
that the $(n+1)$-dimensional bulk, $\mathcal{M}$ is split into two parts by
a hypersurface $\Sigma $, whose two sides will be denoted by $\Sigma _{\pm }$%
. Their normal unit vector, $n^{\mu }$, will be taken to point away from the
surface and into the adjacent space. The gravitational action for the
spacetime $(\mathcal{M}$, $g_{\mu \nu })$ with boundary $(\Sigma $, $h_{ab})$
can be written as \cite{Dav,Mye}
\begin{equation}
I_{G}=\frac{1}{2\kappa ^{2}}\int_{\mathcal{M}}dx^{n+1}\sqrt{-g}\mathcal{L}%
_{G}-\frac{1}{\kappa ^{2}}\int_{\Sigma _{\pm }}dx^{n}\sqrt{-h}\left[
K+2\alpha \left( J-2G^{ab}K_{ab}\right) \right] ,
\end{equation}
where $K_{\mu \nu }$ is the extrinsic curvature of the hypersurface $\Sigma $
with induced metric $h_{ab}$, $G^{ab}$ is the Einstein tensor of the metric $%
h_{ab}$ and $J$ is the trace of
\begin{equation}
J_{ab}=\frac{1}{3}\left( 2KK_{ac}K_{\text{ \ }%
b}^{c}+K_{cd}K^{cd}K_{ab}-2K_{ac}K^{cd}K_{db}-K^{2}K_{ab}\right) .
\end{equation}
The first term in Eq. (\ref{Actg}) is the Einstein-Gauss-Bonnet terms, and
the second term is the boundary term which is chosen such that the
variational principle is well-defined \cite{Mye}. This term is, in fact, the
generalization of Gibbons-Hawking boundary term in Einstein gravity. If we
also include a matter contribution to the action
\begin{equation}
I_{\mathrm{mat}}=-\int_{\mathcal{M}}dx^{n+1}\sqrt{-g}\mathcal{L}_{m}^{(%
\mathcal{M})}-\int_{\Sigma }dx^{n}\sqrt{-h}\mathcal{L}_{m}^{(\Sigma )},
\end{equation}
then the variation of the total action $I=I_{G}+I_{\mathrm{mat}}$ over the
metric tensor $g_{\mu \nu }$ gives\ the equation of gravitational fields in
the bulk as
\begin{eqnarray}
&&\ \mathcal{R}_{\mu \nu }-\frac{1}{2}g_{\mu \nu }\mathcal{R}-\alpha \Bigg\{\frac{%
1}{2}g_{\mu \nu }(\mathcal{R}_{\kappa \lambda \rho \sigma }\mathcal{R}%
^{\kappa \lambda \rho \sigma }-4\mathcal{R}_{\rho \sigma }\mathcal{R}^{\rho
\sigma }+\mathcal{R}^{2})  \nonumber \\
&&-2\mathcal{RR}_{\mu \nu }+4\mathcal{R}_{\mu \lambda
}\mathcal{R}_{\text{ \ }\nu }^{\lambda }+4\mathcal{R}^{\rho \sigma
}\mathcal{R}_{\mu \rho \nu \sigma }-2\mathcal{R}_{\mu }^{\ \rho
\sigma \lambda }\mathcal{R}_{\nu \rho \sigma \lambda
}\Bigg\}=\kappa ^{2}\mathcal{T}_{\mu \nu },  \label{Geq}
\end{eqnarray}
where $\mathcal{T}_{\mu \nu }$ is the energy momentum of the bulk defined by
$\mathcal{T}_{\mu \nu }=$ $2\delta \mathcal{L}_{m}^{(\mathcal{M})}/\delta
g^{\mu \nu }-g_{\mu \nu }\mathcal{L}_{m}^{(\mathcal{M})}$. Assuming the $%
Z_{2}$-symmetry for the brane, then the variation of total action with
respect to $h_{ab}$ gives \cite{Dav}
\begin{equation}
K_{ab}-Kh_{ab}+2\alpha (3J_{ab}-Jh_{ab}+2P_{acbd}K^{cd})=-\frac{\kappa ^{2}}{%
2}T_{ab},  \label{Beq}
\end{equation}
where $T_{ab}=$ $2\delta \mathcal{L}_{m}^{(\Sigma )}/\delta g^{ab}-g_{ab}%
\mathcal{L}_{m}^{(\Sigma )}$ is the energy momentum in the hypersurface and $%
P_{acbd}$ is the divergence free part of the Riemann tensor of the metric $%
h_{ab}$%
\begin{equation}
P_{acbd}=R_{acbd}+2h_{a[c}R_{d]b}+2h_{b[d}R_{c]a}+Rh_{a[c}h_{d]b}.
\end{equation}
Also it is worthwhile to mention that the energy-momentum conservation on
the hypersurface may be written as
\begin{equation}
D^{b}T_{ab}=-2\mathcal{T}_{\mu \nu }h_{a}^{\text{ \ }\mu }n^{\nu },
\label{enc}
\end{equation}
where $D^{b}$ denotes the covariant derivative corresponding to $h_{ab}$.

\section{The de Sitter solutions in an empty bulk}

\label{dS}

\label{Lon}Here we want to obtain the $(n+1)$-dimensional solutions of Eq. (%
\ref{Geq}) in vacuum, which generates a de sitter spacetime. We assume that
the $(n+1)$-dimensional spacetime be isotropic and homogenous, i.e. it has a
maximally symmetric $n$-dimensional space. Since we are interested in
cosmological solutions, we take a metric of the form
\begin{equation}
ds^{2}=-dt^{2}+f^{2}(t)\left[ \frac{dr^{2}}{1-kr^{2}}+r^{2}d\Omega ^{2}%
\right] ,  \label{Met1}
\end{equation}
where $d\Omega ^{2}$ is the metric of an $(n-1)$-sphere and $k$ is a
constant which can take only the value $1$, $0$ and $-1$. To find the
function $F(t)$, one may use any components of Eq. (\ref{Geq}). The simplest
equation is the $tt$-component of these equations which can be written as
\begin{equation}
(\dot{f}^{2}+k)\left[|\alpha| (n-2)(n-3) (\dot{f}^{2}+k)-f^{2}\right]=0 ,
\label{ttcomp}
\end{equation}
where the overdot denotes a derivative with respect to the $t$ coordinate.
The solutions of Eq. (\ref{ttcomp}) can be written as
\begin{equation}
f(t)=\frac{1}{2\sqrt{AB}}\left\{ A\exp \left( \frac{t}{\sqrt{%
(n-2)(n-3)|\alpha| }}\right) +Bk|\alpha| \exp \left( -\frac{t}{\sqrt{%
(n-2)(n-3)|\alpha| }}\right) \right\} ,  \label{Ft}
\end{equation}
where $A$ and$\ B$ are two arbitrary functions. Also for negative $k$, $f(t)=%
\sqrt{-k}t$ is another solution. The function $f(t)$ given in Eq. (\ref{Ft})
shows that the $(n+1)$-dimensional bulk spacetime has an expanding feature.
For example, if one choose a brane in the hypersurface of $r=r_{0}$, then
the volume of $(n-1)$-brane is
\begin{equation}
V_{n-1}=\frac{2\pi ^{n/2}}{\Gamma (n/2)}f^{n-1}(t).  \label{Pot1b}
\end{equation}
It is worthwhile to note that the function $f(t)$ has exponential behavior
for the case of flat space with $k=0$. That is
\begin{equation}
f(t)=A\exp \left( \frac{t}{\sqrt{(n-2)(n-3)|\alpha| }}\right) .
\end{equation}
Also, one may note that the spacetime is asymptotically de Sitter
for an arbitrary value of $k$. In these two cases ($k=0$ and
$t\rightarrow\infty$) the curvature of the spacetime is
$n(n+1)/\{(n-2)(n-3)|\alpha|\}$, which means that the spacetime is
de Sitter.

\section{Modified Friedmann equations in Gauss-Bonnet gravity}

\label{Fried}

Now we consider a 5-D bulk spacetime with a single 4-D hypersurface at $%
r=r_{0}$. The most general gravitational equation in five dimension which is
symmetric, divergence free and linear in second derivative of the metric is
Einstein-Gauss-Bonnet equations. Thus, we use these equations in order to
consider the 4-D brane world cosmology. The brane world $(\mathcal{B},$ $%
h_{\mu \nu })$ is located at the hypersurface $r=r_{0}$, and its induced
metric is $h_{\mu \nu }=g_{\mu \nu }-n_{\mu }n_{\nu }$, where $n^{\mu }=(0,%
\sqrt{g_{rr}},0,0,0)$ is the normal unit vector to the brane. Since we
assume to have a perfect fluid in the brane world at $r=r_{0}$, the bulk has
a spherical symmetric 4-D space. Thus, the bulk metric can be written as:
\begin{equation}
ds^{2}=-dt^{2}+f^{2}(t)\left[ \frac{dr^{2}}{c^{2}(r)}+d\chi ^{2}+b^{2}(\chi
)(d\phi ^{2}+\sin ^{2}\phi d\psi ^{2})\right] ,  \label{Metr2}
\end{equation}
where $b(\chi )$ for the closed, flat and open Universe is
\begin{equation}
b(\chi )=\left\{
\begin{array}{ll}
\sin \chi  & k=+1; \\
\chi  & k=0; \\
\sinh \chi \;\; & k=-1;
\end{array}
\right.
\end{equation}
respectively. The induce metric is
\begin{equation}
ds^{2}=-dt^{2}+a^{2}(t)\left[ d\chi ^{2}+b^{2}(\chi )(d\phi
^{2}+\sin ^{2}\phi d\psi ^{2})\right] ,  \label{Metr3}
\end{equation}
where $a(t)=r_{0}f(t)$. The nonvanishing components of the extrinsic
curvature of the hypersurface are
\begin{eqnarray}
K_{\chi \chi } &=&c_{0}a(t),  \nonumber \\
K_{\phi \phi } &=&c_{0}a(t)b^{2}(\chi ),  \nonumber \\
K_{\psi \psi } &=&c_{0}a(t)b^{2}(\chi )\sin ^{2}\phi ,  \label{ext}
\end{eqnarray}
where $c_{0}=c(r_{0})$. Using Eqs. (\ref{Beq}) and (\ref{ext}), and the
expression of the energy-momentum tensor of a perfect fluid
\begin{equation}
T_{ab}=(\rho +p)u_{a}u_{b}+ph_{ab},  \label{ener}
\end{equation}
one obtains the following modified Friedmann equations:
\begin{eqnarray}
&&\frac{\ddot{a}}{a}=\frac{4c_{0}-\kappa ^{2}pa}{16c_{0}|\alpha |},
\label{acc} \\
&&24c_{0}|\alpha |(k+\dot{a}^{2})+16c_{0}^{3}|\alpha |-6c_{0}a^{2}-\kappa
^{2}\rho a^{3}=0.  \label{Fr2}
\end{eqnarray}
Equation (\ref{acc}) indicates that in matter-dominated era ($p=0$), our
Universe expands with acceleration instead of deceleration along the scheme
of standard Friedmann model. To be more clear we obtain a class of solutions
of Eqs. (\ref{acc}) and (\ref{Fr2}). First, we go through the conservation
of energy-momentum. Using Eqs. (\ref{enc}) and (\ref{ener}) for the metric (%
\ref{Metr2}), and the fact that the bulk is empty, one obtains
\begin{equation}
d(\rho a^{3})=-pda^{3}.  \label{ec}
\end{equation}
Given an equation of state $p=p(\rho )$, one can use Eq. (\ref{ec}) to
determine $\rho $ as a function of $a$. Knowing $\rho $ as a function of $a$%
, one can find $a(t)$ for all time by solving Eq. (\ref{Fr2}). Incidentally,
the solution $a(t)$ determined in this way will automatically satisfy Eq. (%
\ref{acc}), for by differentiating Eq. (\ref{Fr2}) with respect to time and
using Eq. (\ref{ec}), one obtains Eq. (\ref{acc}).

\subsection{Matter dominated era}

Assuming the energy density of Universe is dominated by non-relativistic
matter with negligible pressure, then $\rho \varpropto a^{-3}$, and
therefore Eq.(\ref{Fr2}) can be written as
\begin{equation}
24c_{0}|\alpha |(k+\dot{a}^{2})-6c_{0}a^{2}+16c_{0}^{3}|\alpha |-\kappa
^{2}\rho _{0}a_{0}^{3}=0,  \label{mdeq}
\end{equation}
where $\rho _{0}$ and $a_{0}$ are the values of $\rho $ and $a$ at the
present. Equation (\ref{mdeq}) can be solved exactly. One can write the
solutions of this equation as
\begin{equation}
a(t)=A\exp \left( \frac{t}{2\sqrt{|\alpha |}}\right) +B_{k}\exp \left( -%
\frac{t}{2\sqrt{|\alpha |}}\right) .  \label{At}
\end{equation}
where $B_{k}=(16c_{0}^{3}|\alpha |+24|\alpha |kc_{0}-\kappa ^{2}\rho
_{0}a_{0}^{3})/(24A)$. Since $c_{0}$ is arbitrary, therefore $A$ and $B_{k}$
are two arbitrary constants. The arbitrary constants $A$ and $B_{k}$, and
the Gauss-Bonnet parameter $\alpha $ may be fixed by observational data. In
order to give the way of fixing $A$, $B_{k}$ and $\alpha $, it is more
convenient to work with the Hubble $(H\equiv a^{-1}\dot{a})$, deceleration $%
(q\equiv -H^{-2}a^{-1}\ddot{a})$, jerk $(j\equiv H^{-3}a^{-1}a\stackrel{...}{%
a})$, snap $(s\equiv H^{-4}a^{-1}\stackrel{....}{a})$ and ... parameters
\cite{Vis}. It is remarkable to note that for $a(t)$ given in Eq. (\ref{At}%
), one encounters only with Hubble and deceleration parameters
($H_0$ and $q_0$) at present epoch. Indeed, for $a(t)$ given in
Eq. (\ref{At}) $d^n a/dt^n=H^n q^{[n/2]}$, and therefore the
Taylor expansion of $a(t)$ is
\begin{equation}
a(t)=a_0\sum_{n=0}^{\infty}  {1\over n!}{H_0}^n
{|q_0|}^{[n/2]}(t-t_0)^n,
\end{equation}
where $[n/2]$ denotes the integer part of $n/2$.

In order to use the astronomical data for fixing $(A$, $B_{k}$,
$\alpha )$ or equivalently $(a_{0}$, $H_{0}$, $q_{0})$, one may
use the relation between the luminosity distance $d_{L}$ and the
redshift $z$ of a luminous source. It is easy to show that (see
the Appendix):
\begin{eqnarray}
d_{L} &=&\frac{z}{H_{0}}\Bigg\{1+\frac{1}{2}(1-q_{0})z+\frac{1}{6}%
\left[1-2q_{0}-3q_{0}^{2}+\frac{k}{H_{0}^{2}a_{0}^{2}}\right]z^{2} \nonumber \\
&&\hspace{1.2cm}+\frac{1}{24}\left[1-7q_{0}-24q_{0}^{2}-15q_{0}^{3}+\frac{2k(1+q_{0})}{%
H_{0}^{2}a_{0}^{2}}\right]z^{3}+...\Bigg\}
\end{eqnarray}
Having enough astronomical data, one can determines $(a_{0}$, $H_{0}$, $%
q_{0})$ and therefore $(A$, $B_{k}$, $\alpha )$. The solution (\ref{At})
indicates that during the matter-dominated era, the Universe expands with
acceleration instead of deceleration with positive jerk. This feature is
consistent with the observation of high red-shift supernova or the
measurement of angular fluctuations of cosmic microwave background
fluctuations \cite{Per,Lee,Ton}. Also, one may note that for $a(t)$ given in
Eq. (\ref{At}) $H_{\infty }=(2\sqrt{|\alpha |})^{-1}$ and $s_{\infty
}=j_{\infty }=-q_{\infty }=1$ as $t\rightarrow \infty $.

\section{CLOSING REMARKS}

In this paper, we added the Gauss-Bonnet term with negative coefficient to
the Einstein action without a cosmological constant term, and introduced a
solution of the field equations in an empty $(n+1)$-dimensional bulk. We
found that this solution is de Sitter with constant curvature $%
n(n+1)/2|\alpha |$ for the case of flat space ($k=0$) or as $t\rightarrow
\infty $. We showed that an $(n-1)$-dimensional brane embedded in the bulk
can have an expanding feature with acceleration instead of deceleration.

We also considered the 4-dimensional brane world in a
5-dimensional empty space with zero cosmological constant. This
was done in Gauss-Bonnet gravity which has the most general
gravitational field equation in five dimension. We obtained the
modified Friedmann equations and showed that these two equations
are not functionally independent if one use the conservation of
energy. Equation (\ref{acc}) indicates that our Universe expands
with acceleration instead of deceleration in matter-dominated era.
This was done without need to any mysterious fluid with large
negative pressure or the cosmological constant. We also obtained
the solutions of the modified Friedmann equations for the
matter-dominated era. This solution presents an expanding Universe
with positive acceleration and jerk which is consistent with the
recent cosmological data. We also found that $a(t)$ can be
expressed in terms of $a_0$, $H_0$ and $q_0$ only, and all the
other parameters such as jerk, snap and so on can be written in
terms of $H_0$ and $q_0$.

As stated before, the Gauss-Bonnet gravity is the most general
gravitational field equation in five dimension. In higher
dimension one should use more terms for action in Lovelock theory.
The application of the above method in higher dimension, which
needs the use of Lovelock gravity with more gravitational terms
remains to be carried out in future.

\begin{center}
{\bf{APPENDIX}}
\end{center}
In this appendix, we obtain the relation between the luminosity distance $%
d_{L}$ and the redshift $z$. This relation has been derived for
Robertson-Walker spacetime in \cite{Vis}. Here we obtain it for the metric (%
\ref{Metr3}). The physical distance travelled by a photon emitted at time $%
t_{\ast }$ and absorbed at the current epoch $t_{0}$ in term of
$z$ is \cite {Vis}

\begin{eqnarray}
D &=&{\frac{z}{H_{0}}}\Bigg\{1-\left[ 1+{\frac{q_{0}}{2}}\right] {z}+%
\left[ 1+q_{0}+{\frac{q_{0}^{2}}{2}}-{\frac{j_{0}}{6}}\right] z^{2}
\nonumber \\
&&-\left[ 1+{\frac{3}{2}}q_{0}(1+q_{0})+{\frac{5}{8}}q_{0}^{3}-{\frac{1}{2}}%
j_{0}-{\frac{5}{12}}q_{0}j_{0}-{\frac{s_{0}}{24}}\right] z^{3}+O(z^{4})%
\Bigg\},  \label{Dz}
\end{eqnarray}
while $d_{L}$ can be written as
\begin{equation}
d_{L} =a(t_{0})^{2}\;{\frac{r_{0}}{a(t_{\ast })}}={\frac{a_{0}^{2}}{%
a(t_{0}-D)}b}(\chi )
={\frac{a_{0}^{2}}{a(t_{0}-D)}\{1-}\frac{1}{6}\chi ^{3}+O(\chi
^{5})\}. \label{dL}
\end{equation}
Recall that for a null geodesic in the spacetime (\ref{Metr3})
\begin{eqnarray}
\chi _{0} &=&\int_{t_{\ast }}^{t_{0}}{\frac{dt}{a(t)}}%
=\int_{t_{0}-D}^{t_{0}}{\frac{cdt}{a_{0}}}\;\Bigg\{1+H_{0}(t_{0}-t)+\left[
{\frac{2+q_{0}}{2}}H_{0}^{2}\right] (t_{0}-t)^{2}  \nonumber \\
&&\hspace{3cm}+\left[ {\frac{6(1+q_{0})+j_{0}}{6}}H_{0}^{3}\right]
(t_{0}-t)^{3}+O[(t_{0}-t)^{4}]\Bigg\}  \nonumber \\
&&  \nonumber \\
&&={\frac{D}{a_0}}\Bigg\{1+{\frac{1}{2}}{H_{0}D}+\left[ {\frac{%
2+q_{0}}{6}}\right] \left( {H_{0}D}\right) ^{2}
\nonumber \\
&&\hspace{1cm} +\left[ {\frac{6(1+q_{0})+j_{0}}{24}}\right] \left(
{H_{0}D}\right) ^{3}+O\left[ \left( H_{0}D\right) ^{4}\right]
\Bigg\}. \label{chi0}
\end{eqnarray}
Now using Eqs. (\ref{Dz})-(\ref{chi0}), one obtains
\begin{eqnarray}
d_{L}(z) &=&{\frac{z}{H_{0}}}\Bigg\{1+{\frac{1}{2}}\left[ 1-q_{0}\right] {%
z}-{\frac{1}{6}}\left[ 1-q_{0}-3q_{0}^{2}+j_{0}+{\frac{kc^{2}}{%
H_{0}^{2}a_{0}^{2}}}\right] z^{2}  \nonumber \\
&&+{\frac{1}{24}}\left[
2-2q_{0}-15q_{0}^{2}-15q_{0}^{3}+5j_{0}+10q_{0}j_{0}+s_{0}+{\frac{%
2kc^{2}(1+3q_{0})}{H_{0}^{2}a_{0}^{2}}}\right] z^{3}  \nonumber \\
&&+O(z^{4})\Bigg\}.  \label{dLz}
\end{eqnarray}

\end{document}